# Theory of High-$T_C$ Superconductivity: Accurate Predictions of $T_C$

**DALE HARSHMAN**

  Physikon Research Corporation, Lynden, WA 98264 USA
  University of Notre Dame, Notre Dame, IN 46556 USA

**ANTHONY FIORY**

  New Jersey Institute of Technology, Newark, NJ 07102 USA

Abstract

The superconducting transition temperatures of high-$T_C$ compounds based on copper, iron, ruthenium and certain organic molecules are discovered to be dependent on bond lengths, ionic valences, and Coulomb coupling between electronic bands in adjacent, spatially separated layers.[1] Optimal transition temperature, denoted as $T_{C0}$, is given by the universal expression $k_B T_{C0} = e^2 \Lambda / \ell \zeta$; $\ell$ is the spacing between interacting charges within the layers, $\zeta$ is the distance between interacting layers and $\Lambda$ is a universal constant, equal to about twice the reduced electron Compton wavelength (suggesting that Compton scattering plays a role in pairing). Non-optimum compounds in which sample degradation is evident typically exhibit $T_C < T_{C0}$. For the 31+ optimum compounds tested, the theoretical and experimental $T_{C0}$ agree statistically to within ± 1.4 K. The elemental high $T_C$ building block comprises two adjacent and spatially separated charge layers; the factor $e^2/\zeta$ arises from Coulomb forces between them. The theoretical charge structure representing a room-temperature superconductor is also presented.

[1] DOI: 10.1088/0953-8984/23/29/295701

## 1. Introduction

The pairing mechanism governing the superconductivity of "high transition temperature ($T_C$) superconductors" has been intensely researched and debated since the discovery of the first superconducting cuprate in 1986. In Refs. [1] and [2], wherein 31+ compounds were considered, it was shown how $T_C$ is defined by just two specific lengths, far fewer parameters than previously suggested.

High-$T_C$ superconductors have layered crystal structures, where $T_C$ depends on bond lengths, ionic valences, and Coulomb coupling between electronic bands in adjacent, spatially separated layers. Analysis of high-$T_C$ materials — augmented to number 36 compounds based on Cu, Fe, Ru and organics [1-3] — has revealed that the optimal transition temperature $T_{C0}$ is given by the universal expression

$$T_{C0} = k_B^{-1} e^2 \Lambda / \ell \zeta . \qquad (1)$$

Here, $\ell$ is the spacing between interacting charges within the layers, $\zeta$ is the distance between interacting layers, $\Lambda$ is a universal constant, equal to about twice the reduced electron Compton wavelength ($\bar{\lambda}_e$), $k_B$ is Boltzmann's constant and $e$ is the elementary charge. Non-optimum compounds in which sample degradation is evident typically exhibit $T_C$ below $T_{C0}$.

## 2. Theoretical calculation

Coulomb interaction forces between adjacent charge reservoir layers determine the transition temperature of high-$T_C$ compounds. The 2D interaction charge density $\ell^{-2}$ is obtained from the expression:

$$\ell^{-2} = \eta\sigma/A, \qquad (2)$$

where

- $\eta$ is the number cuprate planes (or equivalent);
- $\sigma$ is the interacting charge fraction obtained from doping and stoichiometry;
- $A$ is the formula-unit basal-plane area.

The separation between adjacent layers $\zeta$ is obtained by examination of crystal structure.

The length scale $\Lambda$ has been evaluated from experiment [1]:

$$\Lambda = 7.47 \times 10^{-11} \text{ cm} \approx 2\,\lambdabar_e.$$

The interacting charge fraction $\sigma$ is obtained from scaling the doping factor $(x - x_0)$ as

$$\sigma = \gamma\,(x - x_0). \qquad (3)$$

The following charge allocation rules determine the coefficient $\gamma$ [1]:

Rule 1a) — Doping shared equally between two reservoirs produces a factor ½.
Rule 1b) — Sharing between N (typically 2) ions or structural layers produces a factor 1/N.

As an illustrative example, consider the case of the compound $La_{2-x}Sr_xCuO_4$, where, $x = 0.163$, $x_0 = 0$, and application of the above rules yields $\gamma = \tfrac{1}{2} \times \tfrac{1}{2}$. The theoretically calculated $T_{C0} = 37.47$ K compares favorably with the experimental $T_{C0} = 38$ K.

The charge fraction $\sigma$ can alternatively be obtained by scaling to $YBa_2Cu_3O_{6.92}$ according to

$$\sigma = \gamma\,\sigma_0, \qquad (4)$$

where $\sigma_0$ is the charge fraction of $YBa_2Cu_3O_{6.92}$, and a second set of valency scaling rules determine $\gamma$ in equation (4) [1].

## 3. Layered ionic structures

High-$T_C$ superconductors comprise two opposing charge reservoirs, denoted as Type I & Type II, as illustrated for $YBa_2Cu_3O_{7-\delta}$ in Figure 1. Shown are periodicity unit cell $d$ and Coulomb interaction height ζ. Table 1 presents the reservoir structures of the 36 compounds for which results on $T_{C0}$ are presented.

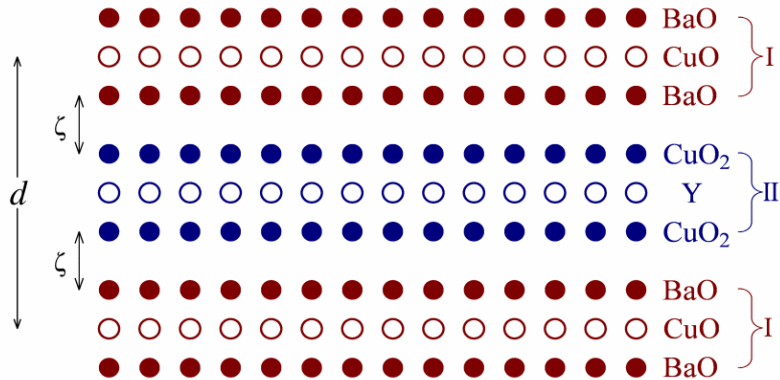

**Figure 1**. Representational cross sectional view of ions in $YBa_2Cu_3O_{7-\delta}$. The three layers in Type I contain Ba and O (filled circles) and Cu and O (open circles). The three layers in Type II contain Cu and O (filled) and Y (open). Height $d$ corresponds to one formula unit per basal-plane area. Distance $\zeta$ is the separation between interacting charge layers.

**Table 1.** Analysis data for 36 optimal high-$T_C$ superconducting compounds (the stoichiometry and hydrostatically applied pressure values correspond to optimal transition temperature $T_{C0}$). Listed are compound formulas, structures of Type I and Type II reservoirs per formula unit ($O_x$ denotes partial filling), and the measured and theoretical $T_{C0}$. Structurally- and doping-related compounds are grouped accordingly (font colors correspond to data point colors in Figures 2 and 3).

| Superconducting Compounds (36) | Type I Reservoirs | Type II Reservoirs | $T_{C0}$ (K) Meas. | $T_{C0}$ (K) Calc. |
|---|---|---|---|---|
| $YBa_2Cu_3O_{6.92}$ | BaO-CuO-BaO | $CuO_2$-Y-$CuO_2$ | 93.7 | 96.36 |
| $YBa_2Cu_3O_{6.60}$ | BaO-CuO-BaO | $CuO_2$-Y-$CuO_2$ | 63 | 64.77 |
| $LaBa_2Cu_3O_{7-\delta}$ | BaO-CuO-BaO | $CuO_2$-La-$CuO_2$ | 97 | 98.00 |
| $YBa_2Cu_4O_8$ (12 GPa) | BaO-CuO-CuO-BaO | $CuO_2$-Y-$CuO_2$ | 104 | 103.19 |
| $Tl_2Ba_2CuO_6$ | BaO-TlO-TlO-BaO | $CuO_2$ | 80 | 79.86 |
| $Tl_2Ba_2CaCu_2O_8$ | BaO-TlO-TlO-BaO | $CuO_2$-Ca-$CuO_2$ | 110 | 108.50 |
| $Tl_2Ba_2Ca_2Cu_3O_{10}$ | BaO-TlO-TlO-BaO | $CuO_2$-Ca-$CuO_2$-Ca-$CuO_2$ | 130 | 130.33 |
| $TlBa_2CaCu_2O_{7-\delta}$ | BaO-TlO-BaO | $CuO_2$-Ca-$CuO_2$ | 103 | 104.93 |
| $TlBa_2Ca_2Cu_3O_{9+\delta}$ | BaO-TlO-BaO | $CuO_2$-Ca-$CuO_2$-Ca-$CuO_2$ | 133.5 | 132.14 |
| $HgBa_2Ca_2Cu_3O_{8+\delta}$ | BaO-$HgO_x$-BaO | $CuO_2$-Ca-$CuO_2$-Ca-$CuO_2$ | 135 | 134.33 |
| $HgBa_2Ca_2Cu_3O_{8+\delta}$ (25 GPa) | BaO-$HgO_x$-BaO | $CuO_2$-Ca-$CuO_2$-Ca-$CuO_2$ | 145 | 144.51 |
| $HgBa_2CuO_{4.15}$ | BaO-$HgO_x$-BaO | $CuO_2$ | 95 | 92.16 |
| $HgBa_2CaCu_2O_{6.22}$ | BaO-$HgO_x$-BaO | $CuO_2$-Ca-$CuO_2$ | 127 | 125.84 |
| $La_{1.837}Sr_{0.163}CuO_{4-\delta}$ | La/SrO-La/SrO | $CuO_2$ | 38 | 37.47 |
| $La_{1.8}Sr_{0.2}CaCu_2O_{6\pm\delta}$ | La/SrO-La/SrO | $CuO_2$-Ca-$CuO_2$ | 58 | 58.35 |
| $(Sr_{0.9}La_{0.1})CuO_2$ | Sr/La | $CuO_2$ | 43 | 41.41 |
| $Ba_2YRu_{0.9}Cu_{0.1}O_6$ | BaO | ½($YRu_{0.9}Cu_{0.1}O_4$) | 35 | 32.21 |
| $(Pb_{0.5}Cu_{0.5})Sr_2(Y,Ca)Cu_2O_{7-\delta}$ | SrO-Pb/CuO-SrO | $CuO_2$-Y/Ca-$CuO_2$ | 67 | 67.66 |
| $Bi_2Sr_2CaCu_2O_{8+\delta}$ (unannealed) | SrO-BiO-BiO-SrO | $CuO_2$-Ca-$CuO_2$ | 89 | 86.65 |
| $(Bi,Pb)_2Sr_2Ca_2Cu_3O_{10+\delta}$ | SrO-BiO-BiO-SrO | $CuO_2$-Ca-$CuO_2$-Ca-$CuO_2$ | 112 | 113.02 |
| $Pb_2Sr_2(Y,Ca)Cu_3O_8$ | SrO-PbO-Cu-PbO-SrO | $CuO_2$-Y/Ca-$CuO_2$ | 75 | 76.74 |
| $Bi_2(Sr_{1.6}La_{0.4})CuO_{6+\delta}$ | SrO-BiO-BiO-SrO | $CuO_2$ | 34 | 34.81 |
| $RuSr_2GdCu_2O_8$ | SrO-$RuO_2$-SrO | $CuO_2$-Gd-$CuO_2$ | 50 | 50.28 |
| $La(O_{0.92-y}F_{0.08})FeAs$ | (1/2)(As-2Fe-As) | ½(La-2O/F-La) | 26 | 24.82 |
| $Ce(O_{0.84-y}F_{0.16})FeAs$ | (1/2)(As-2Fe-As) | ½(Ce-2O/F-Ce) | 35 | 37.23 |
| $Tb(O_{0.80-y}F_{0.20})FeAs$ | (1/2)(As-2Fe-As) | ½(Tb-2O/F-Tb) | 45 | 45.67 |
| $Sm(O_{0.65-y}F_{0.35})FeAs$ | (1/2)(As-2Fe-As) | ½(Sm-2O/F-Sm) | 55 | 56.31 |
| $(Sm_{0.7}Th_{0.3})OFeAs$ | (1/2)(As-2Fe-As) | ½(Sm/Th-2O/F-Sm/Th) | 51.5 | 51.94 |
| $(Ba_{0.6}K_{0.4})Fe_2As_2$ | As-2Fe-As | Ba/K | 37 | 36.93 |
| $Ba(Fe_{1.84}Co_{0.16})As_2$ | As-2(Fe/Co)-As | Ba | 22 | 23.54 |
| $FeSe_{0.977}$ (7.5 GPa) | $Se_{0.997}$ | Fe | 36.5 | 36.68 |
| $Fe_{1.03}Se_{0.57}Te_{0.43}$ (2.3 GPa) | $Se_{0.57}$-$Fe_{0.03}$-$Te_{0.43}$ | $Fe_{1.0}$ | 23.3 | 25.65 |
| $K_{0.83}Fe_{1.66}Se_2$ | Se-$Fe_{1.66}$-Se | $K_{0.83}$ | 29.5 | 30.07 |
| $Rb_{0.83}Fe_{1.70}Se_2$ | Se-$Fe_{1.70}$-Se | $Rb_{0.83}$ | 31.5 | 31.78 |
| $Cs_{0.83}Fe_{1.71}Se_2$ | Se-$Fe_{1.71}$-Se | $Cs_{0.83}$ | 28.5 | 29.44 |
| κ–[BEDT-TTF]$_2$Cu[N(CN)$_2$]Br | S-chains [BEDT-TTF]$_2$ | Cu[N(CN)$_2$]Br | 10.5 | 11.61 |

## 4. Comparison with experiment

Results for superconducting transition temperatures (Table 1) are shown in Figure 2: the ordinate scale is theory, abscissa scale is experiment. The diagonal line is the theoretical expression of equation (1).

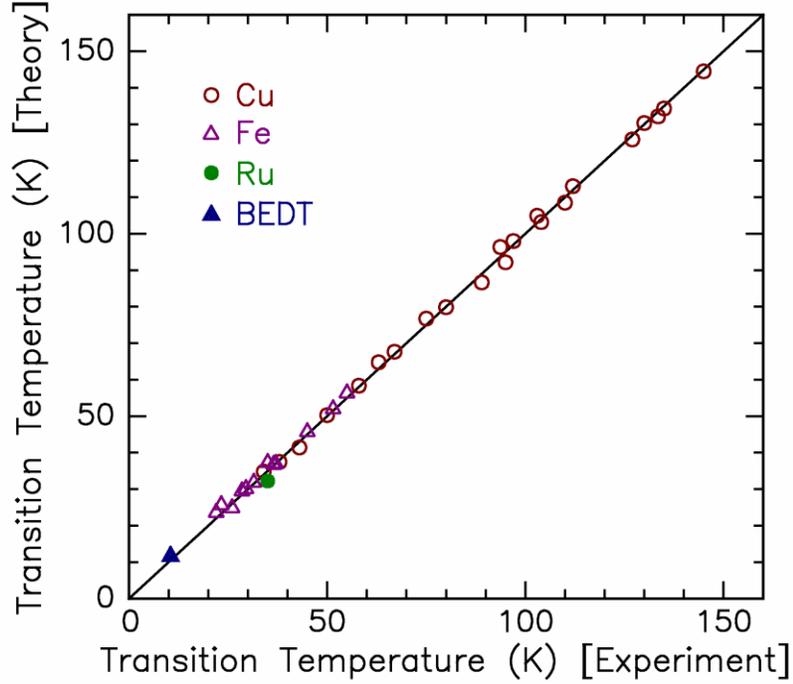

**Figure 2**. Theoretical *vs*. experimentally measured transition temperatures of optimum compounds of Table 1 (colors coordinated), grouped by superconducting compound type: cuprates (Cu), iron pnictides and chalcogenides (Fe), a ruthenate (Ru), an organic (BEDT). Diagonal line is theory, represented by equation (1). Data points fall within ±1.4 K standard deviation of theory.

For superconductivity to occur, the criterion $\zeta \leq \ell$ (interaction distance, $\zeta$, smaller than the mean in-plane separation between interacting charges, $\ell$) is to be obeyed [1]. Figure 3 (left panel) shows the variation of $\zeta$ with experimental transition temperature $T_{C0}$; Figure 3 (right panel) shows the variation of $\zeta$ with experimental transition temperature $T_{C0}$. These results show that in general one finds $\zeta < \ell$.

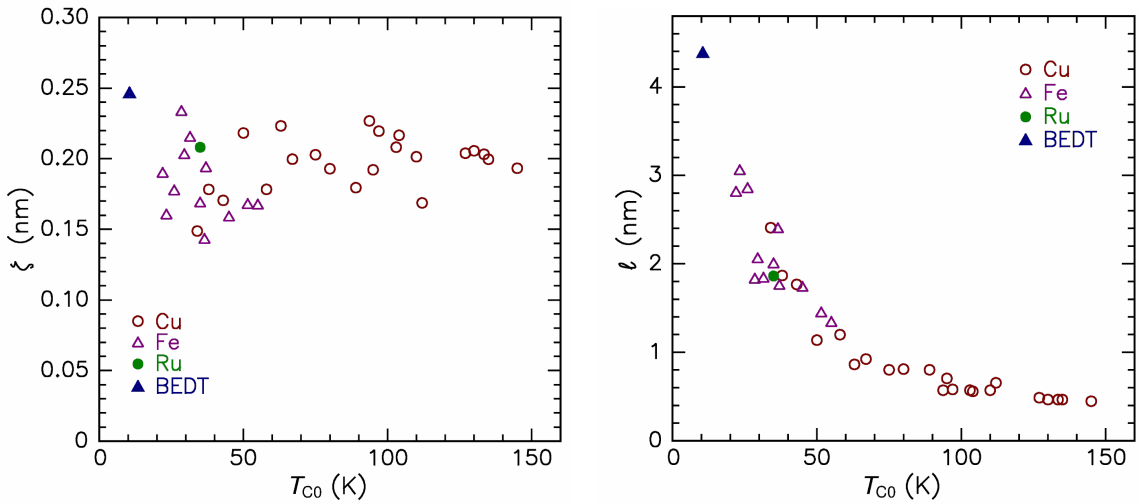

**Figure 3**. Interaction distance $\zeta$ (left panel) and mean in-plane separation between interacting charges $\ell$ (right panel) *vs*. $T_{C0}$ from experiment (compounds listed in Table 1).

## 5. Proposed Superconducting Pairing

The elemental building block of high-$T_C$ superconductors comprises two adjacent and spatially separated charge layers. The factor $e^2/\zeta$ determining $T_{CO}$ arises from Coulomb forces between them. This is illustrated schematically in Figure 4. Two holes (h) form Cooper pair by indirect Coulomb interactions with electronic excitations (e) in the adjacent layer. The superconducting coherence distance $\xi_0$ is generally short compared to mean in-plane separation between interacting charges $\ell$.

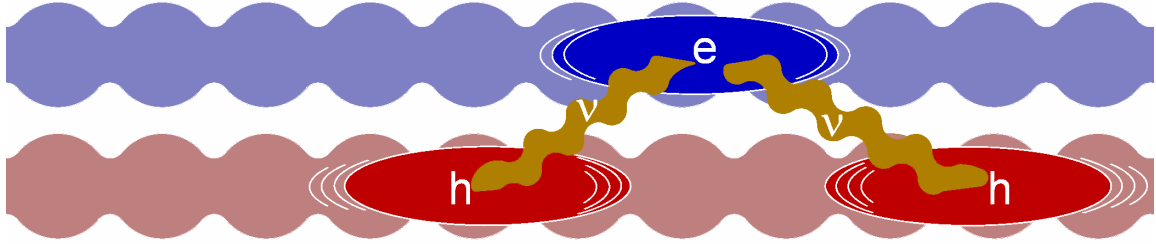

**Figure 4**. Schematic illustration of holes (h) and electrons (e) in spatially separated ionic layers interacting via Coulomb forces, represented by virtual photons (ν).

Remarkably, the theoretical result of equation (1) shows absence of explicit dependence on phonons, plasmons, magnetism, spins, band structure, effective masses, Fermi-surface topologies and pairing-state symmetries in high-$T_C$ materials. The magnitude of $\Lambda$ suggests a universal role of Compton scattering in high-$T_C$ superconductivity, as illustrated in Figure 5, which considers pairing of carriers (h) mediated by electronic excitation (e) via virtual photons (ν).

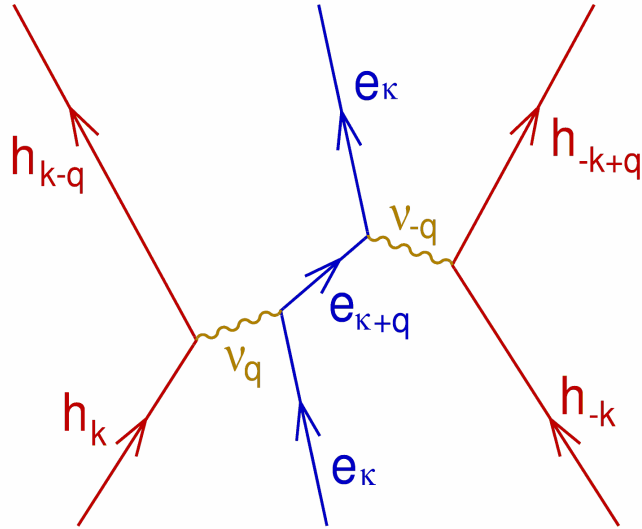

**Figure 5**. Diagram considers pairing of carriers h mediated by electronic excitation e via virtual photons ν.

## 5. Conclusions

1. The theory accurately predicts $T_{C0}$ for the 36 optimal high-$T_C$ superconductors of Table 1, as illustrated in the distribution of errors shown in Figure 6. The mean error between theory and experiment is 1.4 K.

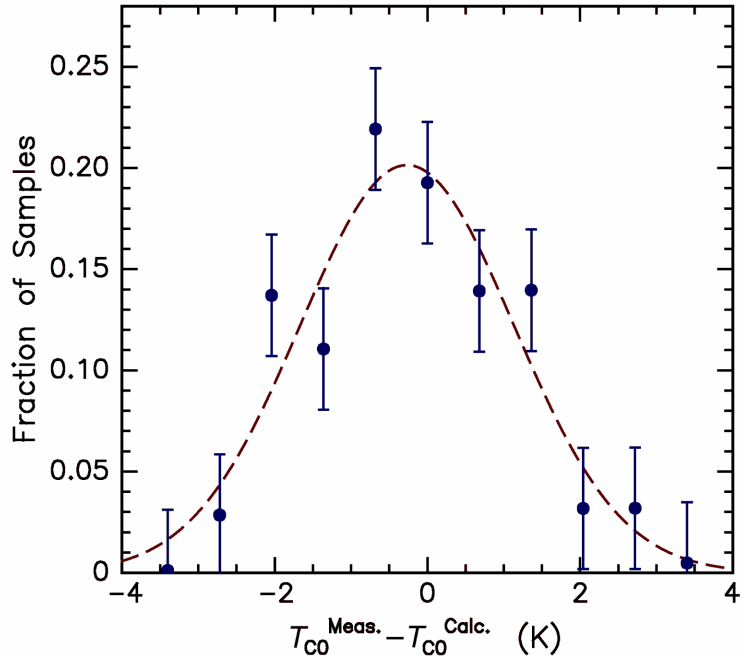

**Figure 6**. Distribution of the deviations between measured and calculated $T_{C0}$. Dashed curve is the normal curve of error of standard deviation 1.4 K (Gaussian function).

2. The minimum superconducting structure contains full Type I and Type II charge reservoirs [2].

3. A conducting charge sheet is non-superconducting without a second mediating charge layer next to it [2].

4. Experimental validations of $\ell$ have been obtained for three cuprates and an organic from measurements of London penetration depths and Hall coefficients, yielding an effective mass $m^* \approx 1.5\, m_0$ and $m^*$ independence of $T_{C0}$ [1]. The length $\ell$ has been verified for the iron-chalcogenide $K_{0.8}Fe_{1.7}Se_2$ from measurements of Fermi wave vector $k_F$ by ARPES [3].

5. The limit $\ell \rightarrow \langle \zeta \rangle$ portends a room temperature superconductor, as equation (1) evaluates to $T_{C0} = 331$ K [1].

6. $T_{C0}$ is proportional to the root-mean-square average of indirect (interlayer) Coulomb forces [1].

This work (which includes supporting information in Refs. 1–4) is contained in a poster presentation at the APS March 2012 meeting [5].

**References**

[1] Theory of high-$T_C$ superconductivity: Transition temperature. D. R. Harshman, A. T. Fiory and J. D. Dow, J. Phys.: Condens. Matter **23** 295701(1-17) (2011). DOI: 10.1088/0953-8984/23/29/295701 [arXiv:1202.0324]

[2] High-$T_C$ superconductivity in ultrathin crystals: Implications for microscopic theory. D. R. Harshman and A. T. Fiory, Emerging Materials Research **1**, 4 (2012). DOI: 10.1680/emr.11.00001 [arXiv:1202.0324]

[3] The Superconducting Transition Temperatures of $Fe_{1+x}Se_{1-y}$, $Fe_{1+x}Se_{1-y}Te_y$ and $(K/Rb/Cs)_zFe_{2-x}Se_2$, D. R. Harshman and A. T. Fiory, J. Phys.: Condens. Matter **24** (2012), *at press* [arXiv:1202.0329]

[4] Coexisting holes and electrons in high-$T_C$ materials: implications from normal state transport. D. R. Harshman, J. D. Dow and A. T. Fiory, Philos. Mag. **91**, 818-840 (2011). DOI: 10.1080/14786435.2010.527864 [arXiv: 1202.1792]

[5] Theory of High-$T_C$ Superconductivity: Accurate Predictions of $T_C$, D. R. Harshman and A. T. Fiory, Bulletin of the American Physical Society, APS March Meeting 2012, Volume 51, Number 1 (Monday–Friday, February 27–March 2, 2012, Boston Massachusetts); Session K1: Poster Session II, Room: Exhibit Hall C, 2:00 PM–2:00 PM, Tuesday, February 28, 2012, Abstract K1.00083.